\documentclass[%
 reprint,
 amsmath,amssymb,
 aps,
 prc,
 floatfix,
]{revtex4-2}

\usepackage{graphicx}
\usepackage{dcolumn}
\usepackage{bm}
\usepackage{nameref}
\usepackage{hyperref}
\usepackage{booktabs}
\usepackage{float}
\usepackage{subcaption}

\begin{document}

\preprint{APS/123-QED}

\title{Investigating the Role of Quantum Entanglement in Heavy Ion Collisions through Elliptic Flow}

\author{Mira Varma}
\email{mira.varma@yale.edu} 
\author{Oliver Baker}
\affiliation{%
 Department of Physics, Yale University, New Haven, CT 06520
}%

\date{\today}

\begin{abstract}
This paper investigates the relationship between initial spatial anisotropy and final state momentum anisotropy in heavy ion collisions through the analysis of elliptic flow ($v_2$) as a function of transverse momentum ($p_T$). Building upon previous studies on thermalization in heavy ion collisions using transverse momentum distributions, we extend the analysis to the $p_T$ dependence of $v_2$ in Pb-Pb and Xe-Xe collisions. By employing a two-component model to extract the thermal and hard scattering contributions to the elliptic flow, we aim to gain further insights into the role of quantum entanglement in the rapid thermalization and collective behavior of the quark-gluon plasma (QGP).
\end{abstract}

\maketitle

\section{Introduction and Definitions}
\begin{figure}[h!]
    \centering
    \includegraphics[trim=5pt 90pt 5pt 90pt, clip, width=0.4\textwidth]{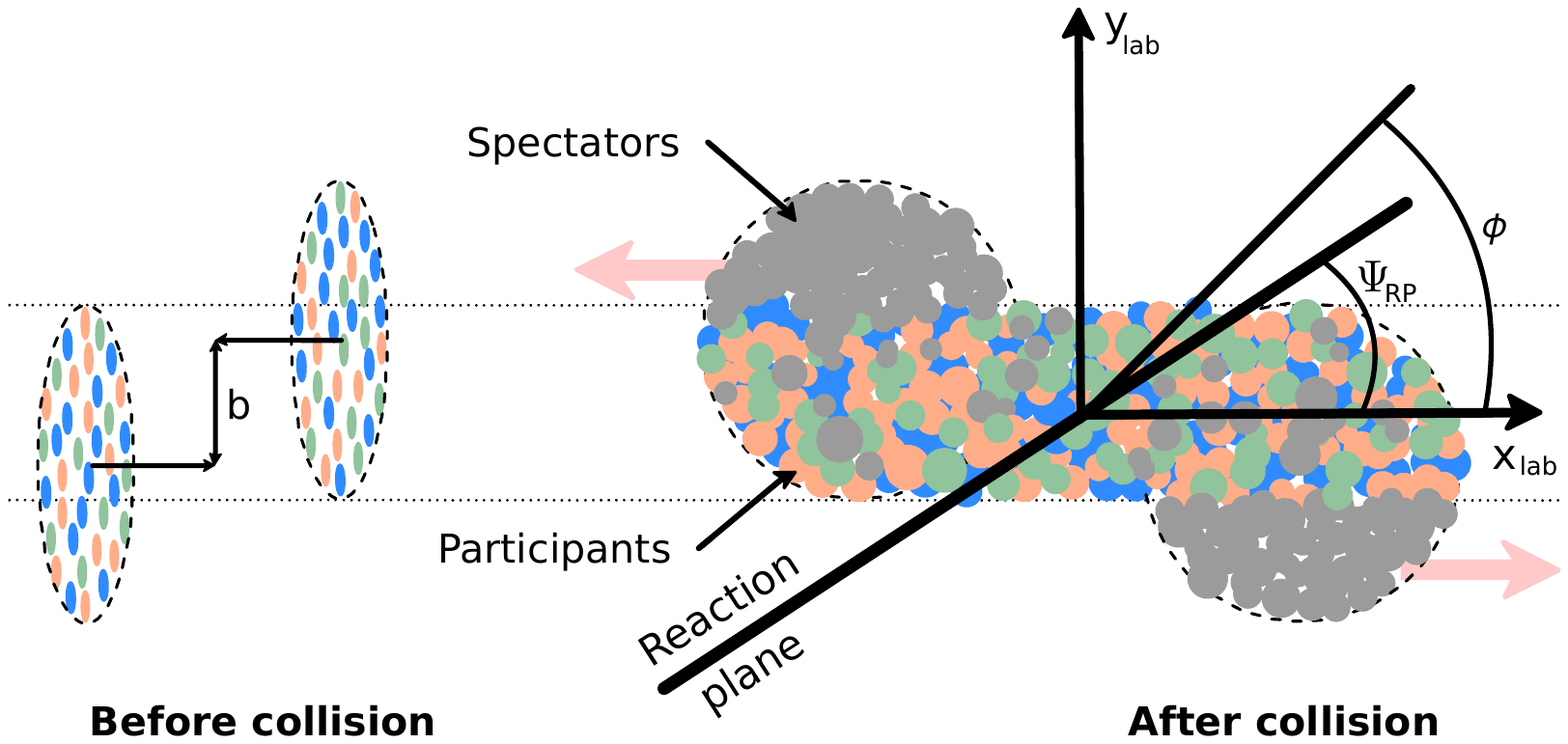}
    \caption{Two heavy ions with non-zero centrality, just prior to collision, characterized by an impact parameter \textbf{b}. Just after the collision, the participants in the overlap region collide to produce multiparticle states that hadronize and are detected in the experimental apparatus. The spectators do not participate in the overlap region collisions. Figure adapted from~\cite{STAR:2014uiw} and~\cite{LHCb:2021ysy}}
    \label{fig:heavy_ions}
\end{figure}

Flow is an experimental observable that provides insights into the transport properties of quarks and gluons produced in heavy ion and hadron-hadron collisions at high energies~\cite{Ollitrault1992, Voloshin1996}. The clearest signature of collective flow is the azimuthal anisotropy in particle production arising from these collisions~\cite{Alver2010, Teaney2011}. A depiction of this process is shown in Fig.~\ref{fig:heavy_ions}. Hydrodynamic models for describing flow are applicable for a description of the system in terms of macroscopic quantities~\cite{Snellings:2011sz}. In Fig.~\ref{fig:heavy_ions}, the reaction plane is defined by the impact parameter $\mathbf{b}$ and the beam direction. The reaction plane angle is $\Psi_{RP}$, and $\phi$ is the azimuthal angle. 

Anisotropic flow is typically characterized by making use of Fourier coefficients, $v_n$, that are functions of transverse momentum, $p_T$, and rapidity, $y$, of the collisions, averaged over all particles in an event~\cite{Derendarz:2014qga}.
\begin{equation}
v_n(p_T, y) = \langle \cos[n(\phi - \Psi_{RP})] \rangle
\end{equation}
The Fourier decomposition coefficient $v_2$ is known as the elliptic flow.

As proton-proton ($p-p$) collision energies increase, the parton density inside the protons gets correspondingly larger. Then, due to the parton interactions and subsequent fragmentation after the collision, the multiplicity of charged (and neutral) hadrons resulting from the reaction grows accordingly. At sufficiently large multiplicities, such as at midrapidity, hadrons interact strongly enough with each other to yield a dynamical azimuthal angle anisotropy in the production and decay process. And while much of the attention has been focused on heavy ion collisions (HICs), the hydrodynamical approach to understanding this phenomenon also includes proton-proton collision processes. There is, according to these authors, reason to believe that collective behavior exists in the final-state rescattering for $p-p$ collisions at LHC energies~\cite{Prasad:2009bx,dEnterria:2010xip,Chaudhuri:2009yp}. A good review of these topics is provided in Ref.~\cite{Snellings:2011sz}.

In this current study, we explore thermal radiation and elliptic flow in the context of both hydrodynamic models and entanglement entropy. Data from both Pb-Pb as well as Xe-Xe collisions at LHC energies are used.

\section{Elliptic Flow and Entanglement Entropy in Heavy Ion Collisions}
In the hydrodynamical approach to the process just described, elliptic flow is an early-time phenomenon, where the produced particles’ rescattering transfers spatial asymmetry into an anisotropic momentum space distribution. The elliptic flow is therefore sensitive to the degree of thermalization in a large part of the process. Recent studies suggest that the pre-equilibrium evolution of the system may significantly influence this thermalization process~\cite{Kurkela2019}. In hydrodynamic models, there must be local thermal equilibration. It is achieved when the mean-free path of constituents, $\lambda$, is small compared to the full system size $R$. 

\begin{figure}[h!]
    \centering
    \includegraphics[trim=45pt 75pt 45pt 75pt, clip, width=0.3\textwidth]{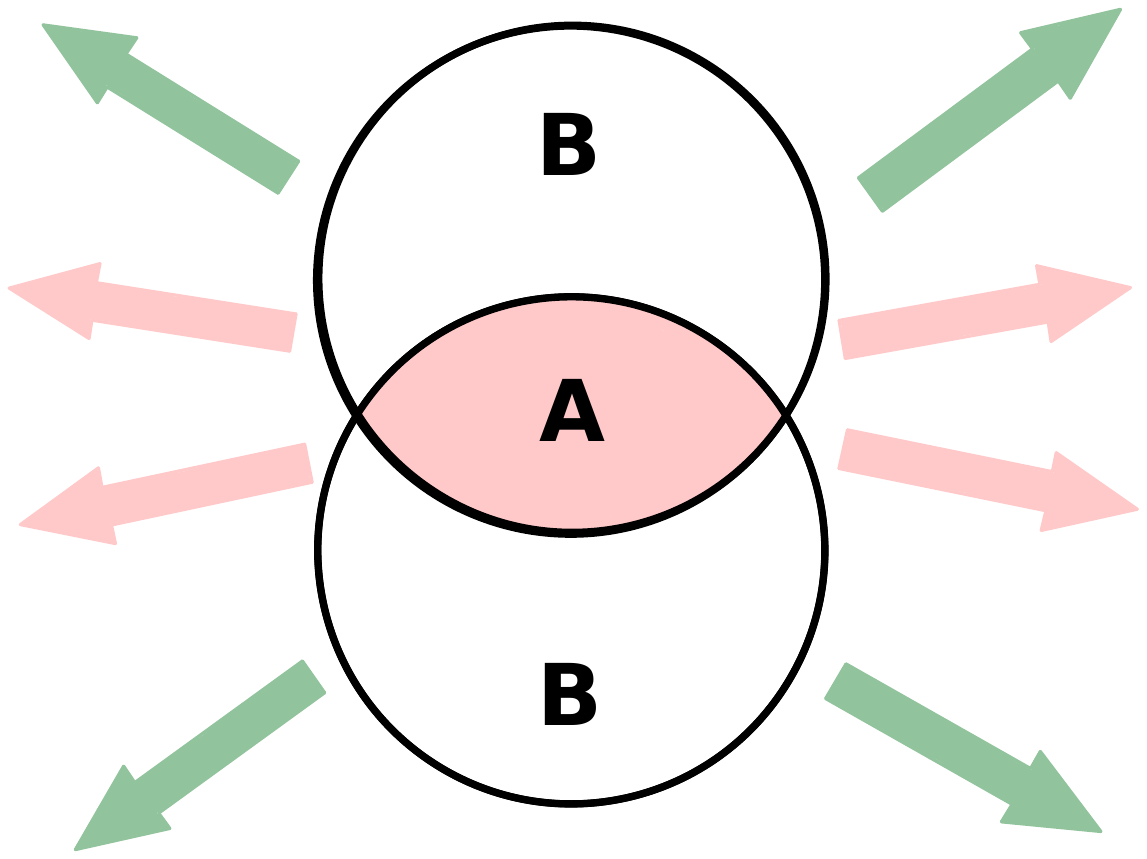}
    \caption{Pictorial description of the overlap (participant) region, $A$, in heavy ion collisions, compared to the non-overlap (spectator) region $B$. The initial nuclei collide at the instant shown here. Multi-particle states emerge from the collision, as indicated by the arrows. Figure adapted from~\cite{Baker2021}.}
    \label{fig:overlap_region}
\end{figure}

Recent studies have shown that quantum entanglement in the initial state wave function can contribute to thermalization~\cite{Baker2018,Ho:2015rga}. In the entanglement entropy picture, we consider region $A$ (the central overlap region) and region $B$ (the spectator region), where the entire space is $A \cup B$. See Fig.~\ref{fig:overlap_region}. Since the nuclei in this interaction are pure states to begin with, regions $A$ and $B$ are entangled. This entanglement entropy gives rise to a thermal component (described by an exponential fit) that dominates over the hard scattering component (described by a power law fit) at low transverse momentum.

To test this quantum entanglement hypothesis quantitatively, we analyze the transverse momentum dependence of the elliptic flow. We analyze the $v_2$ data from Xe-Xe collisions at $\sqrt{s_{NN}} = 5.44$ TeV and Pb-Pb collisions at $\sqrt{s_{NN}} = 5.02$ TeV and 2.76 TeV, measured by the ALICE and ATLAS collaborations~\cite{ALICE:2018lao, Aaboud2018, ALICE2012}. Since the transverse momentum ($p_T$) is much smaller than the longitudinal momenta along the beam direction, the dependence of $v_2$ on $p_T$ can be approximated as~\cite{vanDerKolk2011}:
\begin{equation}
\frac{dv_2(p_T)}{dp_T} \approx \frac{v_2}{\langle p_T \rangle}
\end{equation}

We employ a two-component model to describe $v_2$ as a function of transverse momentum ($p_T$), which allows us to extract the thermal and hard scattering contributions to the elliptic flow~\cite{Zhang2021, Baker2018}.

The thermal contribution is given by:
\begin{equation}
\frac{v_2}{p_T}(\text{thermal}) = A_{\text{th}} \cdot \exp\left(-\frac{m_T}{T_{\text{th}}}\right), \label{thermal_contribution}
\end{equation}
where $A_{\text{th}}$ and $T_{\text{th}}$ are fitting parameters, and $m_T$ is the transverse mass defined as:
\begin{equation}
m_T = \sqrt{m^2 + p_T^2}. \label{transverse_mass}
\end{equation}
In Eq.~\ref{transverse_mass}, we use the pion mass, which does not have a sensitive effect on the data.

The hard scattering contribution is given by:
\begin{equation}
\frac{v_2}{p_T}(\text{hard}) = \frac{A_{\text{hard}}}{\left(1 + \frac{m_T^2}{T^{2n}}\right)^n}, \label{hard_contribution}
\end{equation}
where $A_{\text{hard}}$, $T$, and $n$ are fitting parameters, and $m_T$ is defined as in Eq.~\ref{transverse_mass}

Our two-component model is the sum of the thermal and hard scattering contributions, as described by Equations 4 and 6.

The fits are performed using Python's SciPy package, and the quality of the fits is assessed using $R^2$ (the goodness of fit) and $\chi^2$ values. The ratio of the hard scattering integral to the total fit integral is calculated for each process to quantify the relative contribution of the hard processes to the elliptic flow. The fit parameters, along with their uncertainties, are included in the appendix.

\section{Results and Discussion}

\subsection{Overview of the main findings}
The analysis of the elliptic flow ($v_2$) as a function of transverse momentum ($p_T$) in Xe-Xe collisions at $\sqrt{s_{NN}} = 5.44$ TeV and Pb-Pb collisions at $\sqrt{s_{NN}} = 5.02$ TeV and $2.76$ TeV reveals a complex interplay between thermal and hard scattering processes contributing to the observed anisotropy. The two-component model, which combines thermal and hard scattering contributions, successfully describes the $v_2(p_T)$ data across various centrality classes in all three collision systems. 

In all systems studied, we find: 

\textbf{Thermal Dominance in the Low $\textbf{p}_\textbf{T}$ Region:} In all systems and centralities studied, the thermal component dominates at low $p_T$ ($p_T \lesssim 2$ GeV/c), while the hard scattering component becomes more significant at higher $p_T$. This suggests that the thermal component arises from the collective behavior of the QGP, which is more pronounced at lower $p_T$. The hard scattering component, on the other hand, reflects initial state anisotropies and jet-like correlations, which are more significant at higher $p_T$.

\textbf{Centrality Dependence:} The ratio of the hard scattering integral to the total fit integral exhibits non-monotonic behavior with centrality in Xe-Xe collisions at $\sqrt{s_{NN}} = 5.44$ TeV and Pb-Pb collisions at $\sqrt{s_{NN}} = 5.02$ TeV. In contrast, a monotonic increase is observed in Pb-Pb collisions at $\sqrt{s_{NN}} = 2.76$ TeV. This indicates a sensitivity to the collision conditions.

\textbf{Systematic Trends:} These trends suggest that the interplay between thermal and hard scattering components is a general feature of elliptic flow in heavy-ion collisions. The non-monotonic behavior in certain centrality classes underscores the nuanced dynamics governing the QGP's evolution and the relative contributions of different physical processes across varying collision environments.
This behavior is observed in both Xe-Xe and Pb-Pb collisions, suggesting it is a general feature of elliptic flow in heavy-ion collisions. 
\begin{table}[ht]
\centering
\begin{tabular}{cccc}
\toprule
Centrality & $\frac{\text{hard scattering integral}}{\text{total integral}}$ & $R^2$ & $\chi^2$ \\
\midrule
20-30\% & $0.56 \pm 0.01$ & $0.9959$ & $8.86\times10^{-6}$\\
30-40\% & $0.56 \pm 0.01$ & $0.9952$ & $1.24\times10^{-5}$\\
40-50\% & $0.44 \pm 0.01$ & $0.9954$ & $1.17\times10^{-5}$\\
\bottomrule
\end{tabular}
\caption{Ratios of hard scattering integral to total fit integral for ALICE Xe-Xe collisions at $ \sqrt{s_{NN}} = 5.44 $ TeV, with a pseudorapidity coverage of $|\eta| < 0.8$.}
\end{table}

\begin{table}[ht]
\centering
\begin{tabular}{cccc}
\toprule
Centrality & $\frac{\text{hard scattering integral}}{\text{total integral}}$ & $R^2$ & $\chi^2$ \\
\midrule
20-30\% & $0.44 \pm 0.01$ & $0.9952$ & $6.71\times10^{-6}$\\
30-40\% & $0.60 \pm 0.01$ & $0.9700$ & $5.83\times10^{-6}$\\
40-50\% & $0.50 \pm 0.01$ & $0.9981$ & $4.87\times10^{-6}$\\
\bottomrule
\end{tabular}
\caption{Ratios of hard scattering integral to total fit integral for ATLAS Pb-Pb collisions at $ \sqrt{s_{NN}} = 5.02$ TeV, with a pseudorapidity coverage of $|\eta| < 2.5$.}
\end{table}

\begin{table}[ht]
\centering
\begin{tabular}{cccc}
\toprule
Centrality & $\frac{\text{hard scattering integral}}{\text{total integral}}$ & $R^2$ & $\chi^2$ \\
\midrule
20-30\% & $0.49 \pm 0.01$ & $0.9976$ & $5.76\times10^{-6}$\\
30-40\% & $0.51 \pm 0.01$ & $0.9921$ & $2.15\times10^{-5}$\\
40-50\% & $0.61 \pm 0.01$ & $0.9914$ & $3.52\times10^{-5}$\\
\bottomrule
\end{tabular}
\caption{Ratios of hard scattering integral to total fit integral for ALICE Pb-Pb collisions at $\sqrt{s_{NN}}$ = $2.76$ TeV, with a pseudorapidity coverage of $|\eta| < 0.8$.}
\end{table}

\begin{figure*}[htbp]
    \centering
    \begin{subfigure}[t]{0.4\textwidth}
        \centering
        \includegraphics[width=\textwidth]{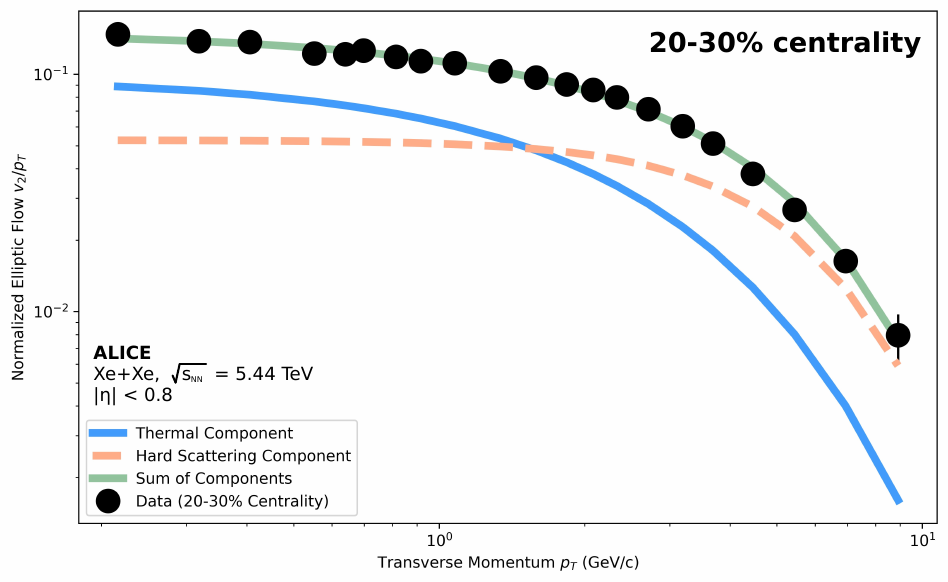}
        \caption{20-30\% centrality}
    \end{subfigure}
    \hfill
    \begin{subfigure}[t]{0.4\textwidth}
        \centering
        \includegraphics[width=\textwidth]{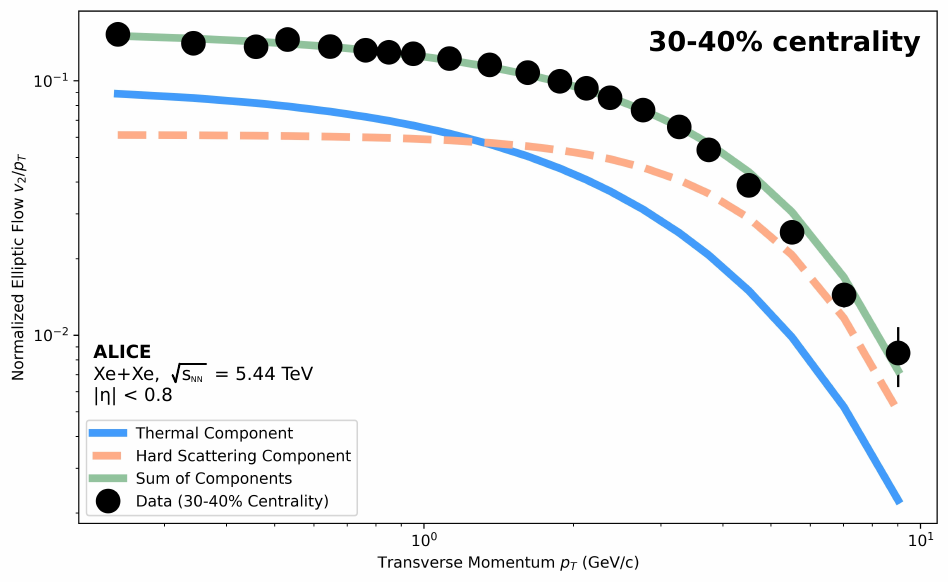}
        \caption{30-40\% centrality}
    \end{subfigure}
    \hfill
    \begin{subfigure}[t]{0.4\textwidth}
        \centering
        \includegraphics[width=\textwidth]{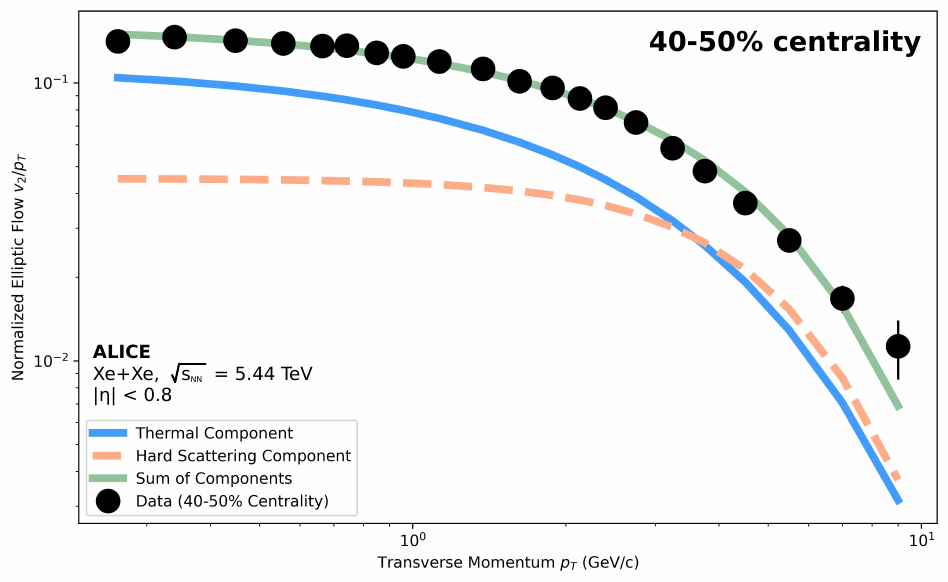}
        \caption{40-50\% centrality}
    \end{subfigure}
    \caption{Elliptic flow $v_2$ vs $p_T$ for Xe-Xe collisions at $\sqrt{s_{NN}} = 5.44$ TeV at various centralities. Data is from~\cite{ALICE:2018lao}.}
    \label{fig:Xe5.44}
\end{figure*}

\begin{figure*}[htbp]
    \centering
    \begin{subfigure}[t]{0.4\textwidth}
        \centering
        \includegraphics[width=\textwidth]{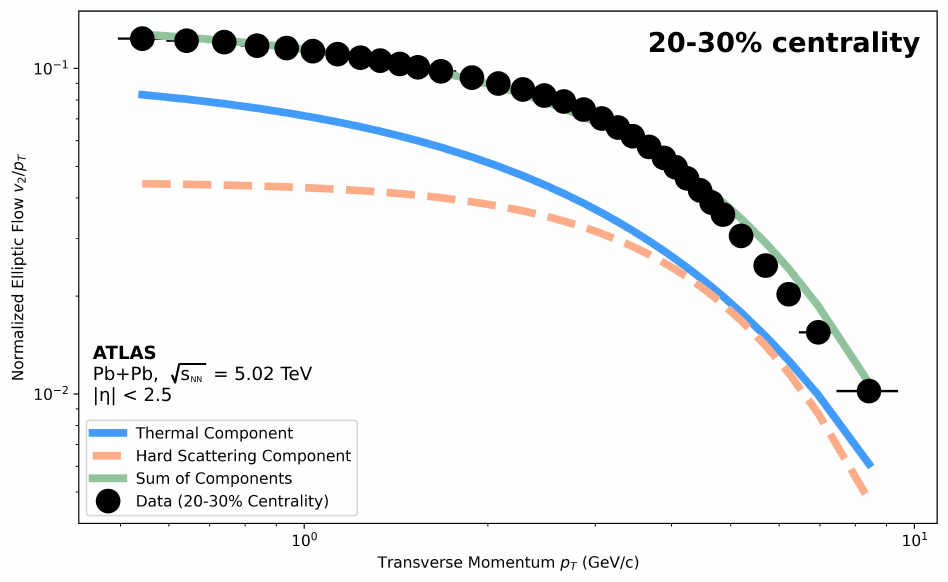}
        \caption{20-30\% centrality}
    \end{subfigure}
    \hfill
    \begin{subfigure}[t]{0.4\textwidth}
        \centering
        \includegraphics[width=\textwidth]{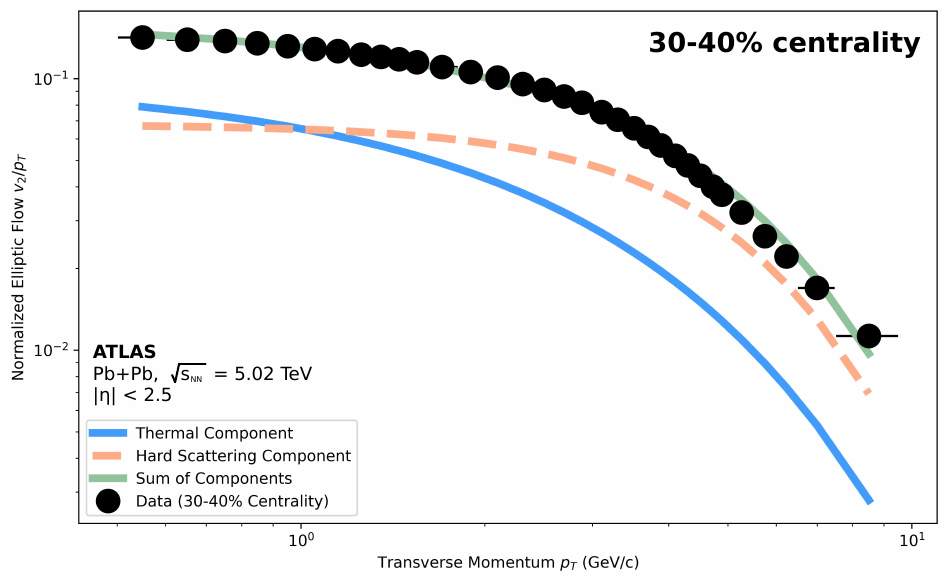}
        \caption{30-40\% centrality}
    \end{subfigure}
    \hfill
    \begin{subfigure}[t]{0.4\textwidth}
        \centering
        \includegraphics[width=\textwidth]{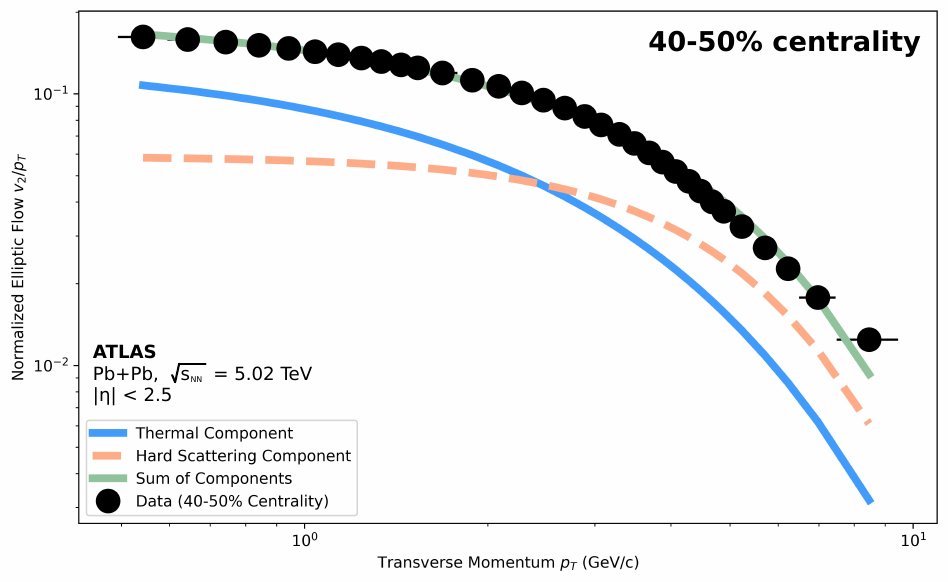}
        \caption{40-50\% centrality}
    \end{subfigure}
    \caption{Elliptic flow $v_2$ vs $p_T$ for Pb-Pb collisions at $\sqrt{s_{NN}} = 5.02$ TeV at various centralities. Data is from~\cite{Aaboud2018}.}
    \label{fig:Pb5.02}
\end{figure*}

\begin{figure*}[htbp]
    \centering
    \begin{subfigure}[t]{0.4\textwidth}
        \centering
        \includegraphics[width=\textwidth]{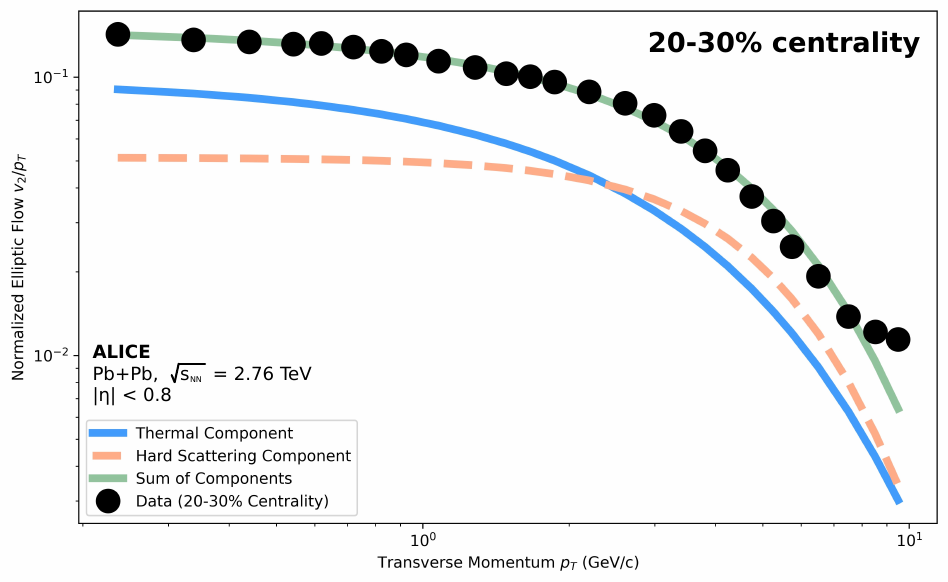}
        \caption{20-30\% centrality}
    \end{subfigure}
    \hfill
    \begin{subfigure}[t]{0.4\textwidth}
        \centering
        \includegraphics[width=\textwidth]{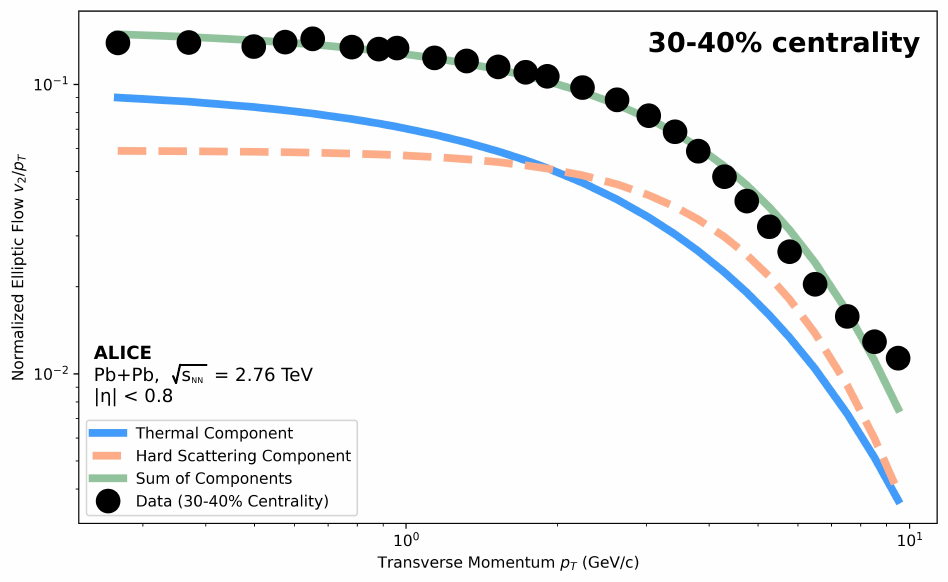}
        \caption{30-40\% centrality}
    \end{subfigure}
    \hfill
    \begin{subfigure}[t]{0.4\textwidth}
        \centering
        \includegraphics[width=\textwidth]{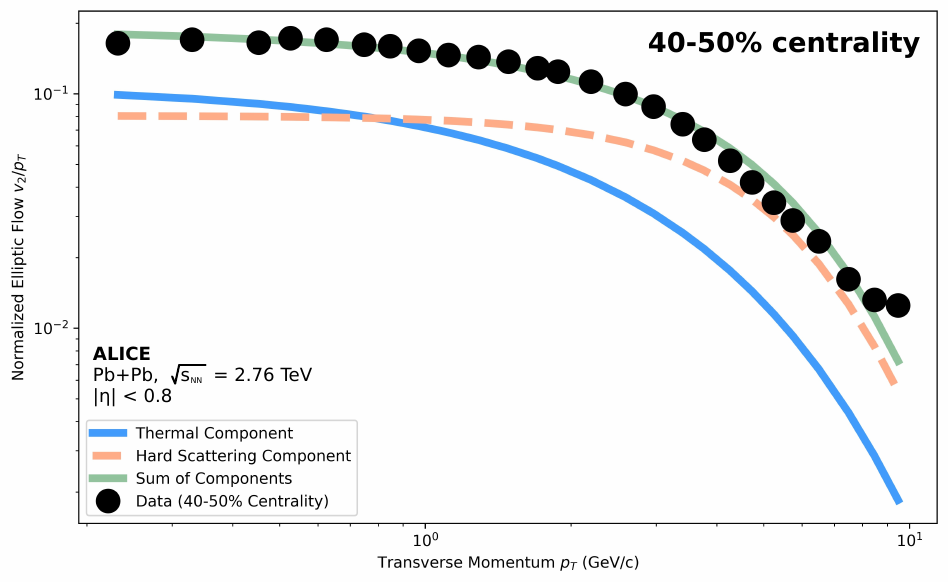}
        \caption{40-50\% centrality}
    \end{subfigure}
    \caption{Elliptic flow $v_2$ vs $p_T$ for Pb-Pb collisions at $\sqrt{s_{NN}} = 2.76$ TeV at various centralities. All error bars are smaller than the data points. Data is from~\cite{ALICE2012}.}
    \label{fig:Pb2.76}
\end{figure*}

\subsubsection{Xe-Xe collisions at $\sqrt{s_{NN}} = 5.44$ TeV}
Xe-Xe collisions at $\sqrt{s_{NN}} = 5.44$ TeV are studied for the 20-30\%, 30-40\%, and 40-50\% centrality classes (Fig.~\ref{fig:Xe5.44}). The two-component model provides a good description of the data in all classes, with high $R^2$ and low $\chi^2$ values (Table I). The thermal component dominates at low $p_T$, transitioning to the hard scattering component at higher $p_T$ ($\sim$ 2-3 GeV/c), depending on centrality. The ratio of the hard scattering integral to the total fit integral shows non-monotonic behavior with centrality, remaining constant between 20-30\% and 30-40\% centrality ($0.56 \pm 0.01$) and decreasing at 40-50\% ($0.44 \pm 0.01$). At low $p_T$ ($<1$ GeV/c), $v_2$ is similar across centralities, indicating consistent collective behavior of the QGP. This similarity is expected from a theoretical perspective, as the low $p_T$ region is dominated by the collective expansion of the QGP, which is less sensitive to the initial state geometry~\cite{Song:2007fn, Noronha-Hostler:2008kkf}.

\subsubsection{Pb-Pb collisions at $\sqrt{s_{NN}} = 5.02$ TeV and $2.76$ TeV}
The $v_2(p_T)$ in Pb-Pb collisions is studied at $\sqrt{s_{NN}} = 5.02$ TeV and 2.76 TeV for the 20-30\%, 30-40\%, and 40-50\% centrality classes (Fig.~\ref{fig:Pb5.02} and~\ref{fig:Pb2.76}). The two-component model accurately describes the data at both energies, with high $R^2$ values and low $\chi^2$ values (Tables II and III). The thermal component dominates at low $p_T$, transitioning to the hard scattering component at higher $p_T$.

Comparing results between the two energies, $v_2$ is generally larger at $\sqrt{s_{NN}} = 5.02$ TeV, particularly at higher $p_T$, likely due to higher energy density and larger QGP volume. The behavior of the hard scattering ratio across centrality classes is similar for both collision energies, with a net increase from 20-30\% to 40-50\% centrality. However, the highest ratio value is observed in the 30-40\% centrality class for $\sqrt{s_{NN}} = 5.02$ TeV ($0.60 \pm 0.01$) but for $\sqrt{s_{NN}} = 2.76$ TeV, the highest ratio value is at 40-50\% centrality ($0.61 \pm 0.01$). At $\sqrt{s_{NN}} = 5.02$ TeV, the hard scattering ratio is lowest at 20-30\% ($0.44 \pm 0.01$), and the 40-50\% centrality class falls in between ($0.50 \pm 0.01$). For $\sqrt{s_{NN}} = 2.76$ TeV, the ratio increases monotonically with centrality, with the lowest value at 20-30\% ($0.49 \pm 0.01$), followed 30-40\% ($0.51 \pm 0.01$), and the highest value at 40-50\% centrality ($0.61 \pm 0.01$).

\subsection{Implications for Possible Quantum Entanglement}
The dominance of the thermal component at low transverse momentum ($p_T \lesssim 2$ GeV/c) across different collision systems provides important insights into possible quantum entanglement effects. Following Ref.~\cite{Ho:2015rga}, in heavy ion collisions, region $A$ (the central overlap region containing mostly soft particles) and region $B$ (containing hard and collinear modes) can become quantum mechanically entangled. When most of the degrees of freedom in $B$ are traced over, the resulting density matrix for region $A$ has very high entropy, which could contribute to the thermal-like behavior we observe. This quantum mechanical mechanism allows thermalization to occur more rapidly than would be possible through conventional kinetic theory requiring multiple sequential scatterings~\cite{Baier2001}.

\subsection{Experimental Considerations}
It is important to note that the ATLAS measurement at 5.02 TeV has a wider pseudorapidity coverage $(|\eta| < 2.5)$ compared to the ALICE measurements at 2.76 TeV $(|\eta| < 0.8)$. This difference in pseudorapidity coverage may influence the observed centrality dependence of the hard scattering ratio, as $v_2$ typically decreases towards forward and backward rapidities~\cite{Voloshin:1994mz}. Additionally, the relative contributions of the thermal and hard scattering processes to the elliptic flow may vary with pseudorapidity~\cite{Poskanzer:1998yz}, which can affect the extracted fit parameters and their centrality dependence.

Another important factor is the variation in data-taking methods employed by the ALICE and ATLAS collaborations. The ALICE collaboration used the event plane method with a pseudorapidity gap of $ |\Delta \eta| > 2.0 $ for the Pb-Pb collisions at $\sqrt{s_{NN}} = 2.76$ TeV and the Xe-Xe collisions at $\sqrt{s_{NN}} = 5.44$ TeV, while the ATLAS collaboration used the scalar product method for the Pb-Pb collisions at $\sqrt{s_{NN}} = 5.02$ TeV. Although both methods aim to suppress non-flow effects, the different techniques may lead to variations in the extracted $v_2$ values and, consequently, the centrality dependence of the hard scattering ratio. Future studies using consistent data-taking methods across different collision systems and energies would help to disentangle the effects of the experimental techniques from the underlying physics.

\section{Conclusion}
The key finding of this study is the non-monotonic behavior of the ratio of the hard scattering integral to the total fit integral with centrality, observed in Xe-Xe collisions at $\sqrt{s_{NN}} = 5.44$ TeV and Pb-Pb collisions at $\sqrt{s_{NN}} = 5.02$ TeV, while a monotonic increase is seen in Pb-Pb collisions at $\sqrt{s_{NN}} = 2.76$ TeV. This behavior reveals a complex interplay between the hard and thermal processes that is sensitive to the collision system and energy. 

By analyzing the elliptic flow ($v_2$) as a function of transverse momentum ($p_T$), we demonstrate that a two-component model is able to describe the Pb-Pb and Xe-Xe collision data across various centrality classes. We find that the thermal component dominates at low $p_T$ ($\lesssim 2$ GeV/c), while the hard scattering component becomes more significant at higher $p_T$. Recent theoretical work suggests that quantum entanglement between different regions of the collision system might play a role in such thermal behavior~\cite{Ho:2015rga}. Establishing this connection requires further theoretical development.

While our study provides valuable insights, it is important to acknowledge the potential limitations of our approach. The two-component model, although successful in describing the elliptic flow data, is a simplified representation of the complex dynamics in heavy ion collisions. Future research could explore more sophisticated models that incorporate quantum effects more explicitly, such as those based on the AdS/CFT correspondence~\cite{Maldacena1999, Gubser2008, Bantilan2012} or the color glass condensate framework~\cite{Gelis2010}. Recent reviews of early-time dynamics~\cite{Schlichting2019} have highlighted how different thermalization mechanisms in weakly coupled non-abelian plasmas may be important during the first fm/c of heavy-ion collisions. Additionally, our analysis focuses on a limited range of collision systems and energies. Extending the study to a wider variety of collision systems, such as proton-nucleus or light ion collisions, and exploring a broader range of energies could provide a more comprehensive understanding of the role of quantum effects in different collision scenarios. Furthermore, the inclusion of additional observables, such as higher-order flow harmonics~\cite{Yan2014} or event-by-event fluctuations~\cite{Giacalone2017}, could offer complementary insights into the interplay between quantum entanglement and the collective behavior of the QGP.

Our analysis reveals systematic patterns in how thermal and hard scattering components contribute to elliptic flow across different collision systems and energies. The persistence of the thermal component under various conditions raises questions about the underlying thermalization mechanisms. While conventional explanations through rescattering remain viable, the possible role of quantum effects, including entanglement, deserves further theoretical and experimental investigation.

\begin{acknowledgments}
The authors gratefully acknowledge funding support from the Department of Energy Office of Science Award DE-FG02-92ER40704.
\end{acknowledgments}

\appendix
\onecolumngrid

\section{Fit Parameters and Uncertainties}
In this appendix, we provide detailed tables of the fit parameters and their uncertainties for the Xe-Xe and Pb-Pb collisions at different energies and centrality classes.

\subsection{ALICE Xe-Xe 5.44 TeV}

\begin{table}[H]
\centering
\begin{tabular}{cccccc}
\toprule
Centrality & $A_{\text{th}}$ & $T_{\text{th}}$ & $A_{\text{hard}}$ & $T$ & $n$ \\
\midrule
20-30\% & $0.1 \pm 2.94\times10^{-6}$ & $2.157 \pm 7.27\times10^{-5}$ & $0.0528 \pm 2.99\times10^{-6}$ & $5.38 \pm 9.91\times10^{-5}$ & $5.0 \pm 1.43\times10^{-3}$ \\
30-40\% & $0.1 \pm 4.44\times10^{-6}$ & $2.376 \pm 1.18\times10^{-4}$ & $0.0614 \pm 4.49\times10^{-6}$ & $5.0 \pm 4.77\times10^{-5}$ & $5.0 \pm 1.12\times10^{-3}$ \\
40-50\% & $0.118 \pm 8.14\times10^{-6}$ & $2.487 \pm 1.87\times10^{-4}$ & $0.0454 \pm 8.21\times10^{-6}$ & $5.0 \pm 1.04\times10^{-4}$ & $5.0 \pm 3.43\times10^{-3}$ \\
\bottomrule
\end{tabular}
\caption{Fit parameters and uncertainties for ALICE Xe-Xe collisions at $\sqrt{s_{NN}} = 5.44$ TeV.}
\end{table}

\subsection{ATLAS Pb-Pb 5.02 TeV}

\begin{table}[H]
\centering
\begin{tabular}{cccccc}
\toprule
Centrality & $A_{\text{th}}$ & $T_{\text{th}}$ & $A_{\text{hard}}$ & $T$ & $n$ \\
\midrule
20-30\% & $0.1 \pm 8.06\times10^{-6}$ & $3.013 \pm 2.68\times10^{-4}$ & $0.0448 \pm 8.16\times10^{-6}$ & $5.0 \pm 7.91\times10^{-5}$ & $5.0 \pm 2.10\times10^{-3}$ \\
30-40\% & $0.1 \pm 2.24\times10^{-6}$ & $2.381 \pm 6.63\times10^{-5}$ & $0.0678 \pm 2.33\times10^{-6}$ & $5.0 \pm 8.02\times10^{-6}$ & $5.0 \pm 2.42\times10^{-4}$ \\
40-50\% & $0.138 \pm 1.78\times10^{-6}$ & $2.249 \pm 3.72\times10^{-5}$ & $0.0589 \pm 1.87\times10^{-6}$ & $5.0 \pm 1.31\times10^{-5}$ & $5.0 \pm 2.15\times10^{-4}$ \\
\bottomrule
\end{tabular}
\caption{Fit parameters and uncertainties for ATLAS Pb-Pb collisions at $\sqrt{s_{NN}} = 5.02$ TeV.}
\end{table}

\subsection{ALICE Pb-Pb 2.76 TeV}

\begin{table}[H]
\centering
\begin{tabular}{cccccc}
\toprule
Centrality & $A_{\text{th}}$ & $T_{\text{th}}$ & $A_{\text{hard}}$ & $T$ & $n$ \\
\midrule
20-30\% & $0.1 \pm 3.31\times10^{-6}$ & $2.712 \pm 1.01\times10^{-4}$ & $0.0515 \pm 3.35\times10^{-6}$ & $5.0 \pm 1.17\times10^{-5}$ & $5.0 \pm 6.00\times10^{-4}$ \\
30-40\% & $0.1 \pm 5.01\times10^{-6}$ & $2.871 \pm 1.59\times10^{-4}$ & $0.0591 \pm 5.06\times10^{-6}$ & $5.0 \pm 2.55\times10^{-5}$ & $5.0 \pm 8.98\times10^{-4}$ \\
40-50\% & $0.111 \pm 2.12\times10^{-6}$ & $2.312 \pm 5.15\times10^{-5}$ & $0.0807 \pm 2.16\times10^{-6}$ & $5.0 \pm 1.41\times10^{-5}$ & $5.0 \pm 2.54\times10^{-4}$ \\
\bottomrule
\end{tabular}
\caption{Fit parameters and uncertainties for ALICE Pb-Pb collisions at $\sqrt{s_{NN}} = 2.76$ TeV.}
\end{table}

\twocolumngrid
\bibliographystyle{unsrt}
\bibliography{my_bib_file}
\end{document}